\documentclass[aps,prd,reprint,superscriptaddress,showpacs,preprintnumbers,floatfix]{revtex4-2}

\usepackage{amsmath,amssymb,amsfonts}
\usepackage{graphicx}
\usepackage{dcolumn}
\usepackage{bm}
\usepackage{slashed}
\usepackage{hyperref}
\hypersetup{colorlinks=true,linkcolor=blue,citecolor=blue,urlcolor=blue}

\begin{document}
	
	\title{Black hole correspondences of quasi-topological gravity:Shadows, quasinormal modes, graybody factors}
	
	\author{Sihao Fan}
	\email{sihaofan02@gmail.com}
	\affiliation{
		University of Shanghai for Science and Technology, Shanghai 200093, China
	}
	
	\author{Chen Wu}
	\email{wuchenoffd@gmail.com}
	\affiliation{Xingzhi College, Zhejiang Normal University, Jinhua 321004, Zhejiang, China}
	\author{Wenjun Guo}
	\affiliation{
		University of Shanghai for Science and Technology, Shanghai 200093, China
	}
	\date{\today}
	
\begin{abstract}
	Within the framework of quasi-topological gravity, we systematically investigate the correspondences among three classes of observables---the shadow, the quasinormal modes, and the graybody factors---of regular black holes in $D = 5$ dimensions. Using the WKB approximation and GrayHawk numerical integration, we test the applicability of the GBF--QNM correspondence and the shadow--GBF correspondence in the low-mode ($l=2$) and higher-dimensions ($D=5$): the difference between the GBF--QNM correspondence and the numerical results remains small in magnitude; we apply the known Langer correction (replacing $l$ by the effective angular momentum $\kappa = \sqrt{l(l+D-3)} \simeq l + (D-3)/2$), which reduces the error of the shadow--GBF correspondence at low modes to a level comparable with that of the GBF--QNM correspondence. Our work extends the shadow--GBF correspondence to higher-dimensional regular black holes, providing a viable route to predict the black hole spectrum and the Hawking radiation profile from a single observable, and thereby to realize multi-messenger tests of gravity.
\end{abstract}
	
	\maketitle
	
	\section{Introduction}
	% ...
	General relativity (GR) predicts that spacetime singularities inevitably form at the endpoint of gravitational collapse, as established by the Penrose singularity theorem under the assumptions of the null convergence condition, global hyperbolicity, and the existence of a closed trapped surface \cite{Hawking1970,Penrose1965}. At a spacetime singularity, curvature invariants diverge and geodesics become incomplete, signaling the breakdown of classical GR and the necessity of new physics beyond it \cite{HawkingEllis1973}. While a complete theory of quantum gravity remains elusive, a productive bottom-up approach has been to construct singularity-free (regular) black hole spacetimes as effective geometries that capture plausible quantum gravitational effects, and to study their observational signatures \cite{Bambi2023,CarballoRubio2025,Lan2023}.
	
	The first regular black hole model was proposed by Bardeen, followed by the Hayward \cite{Hayward2006} and Dymnikova \cite{Dymnikova1992} solutions, among many others \cite{AyonBeato1998,Bonanno2000,Modesto2004,Nicolini2006}. These models share a common feature: a de Sitter-like core replaces the classical singularity at $r=0$, rendering all curvature invariants finite everywhere. However, traditional constructions have a significant limitation---they treat the regular metric as an \textit{ad hoc} ansatz and then reverse-engineer the required matter stress-energy tensor, typically necessitating exotic matter fields such as nonlinear electrodynamics (NED) \cite{AyonBeato1999,Bokulic2022,Canate2023,Dymnikova2004,Fan2016} that often violate energy conditions and require fine-tuning between the theory parameters and integration constants \cite{Zhou2023}.
	
	A major breakthrough was recently achieved by Bueno, Cano, and Hennigar \cite{Bueno2025}, who demonstrated that regular black holes can arise as vacuum solutions of purely gravitational theories in $D \geq 5$ spacetime dimensions, without introducing any matter fields whatsoever. Their construction employs quasi-topological (QT) gravities \cite{Bueno2017,Myers2010,Oliva2010}---a class of higher-curvature theories whose field equations remain second-order on static spherically symmetric backgrounds, guaranteeing uniqueness of bl ack hole solutions and, in many cases, a Birkhoff theorem \cite{Cisterna2017}. The key insight is that an infinite tower of higher-curvature corrections, when resummed under mild convergence conditions on the coupling constants $\alpha_n$, generically resolves the Schwarzschild singularity. The resulting family of exact solutions includes higher-dimensional generalizations of the Hayward, Bardeen, and Dymnikova regular black holes, as well as several novel variants. Crucially, the QT family provides a complete basis for vacuum gravitational effective field theory (EFT) in $D \geq 5$ \cite{Bueno2019}, meaning these regular black holes are not arbitrary constructions but genuinely capture the low-energy limit of whatever quantum gravity theory underlies GR. Establishing purely gravitational singularity resolution in this framework is therefore a landmark result that motivates a systematic investigation of the observational signatures of these regular black holes---a task we undertake in this paper.
	
	Three key observables encode distinct aspects of black hole physics and are, in principle, accessible to current and next-generation instruments. Quasinormal modes (QNMs) \cite{Kokkotas1999,Konoplya2011,Nollert1999} are the characteristic damped oscillations that dominate the ringdown phase of a perturbed black hole; their complex frequencies $\omega_{\mathrm{QNM}} = \omega_{R} + i\,\omega_{I}$ probe the near-horizon geometry and the shape of the effective potential barrier. The detection of multiple QNM overtones in gravitational wave signals from binary black hole mergers \cite{Abbott2016,Abbott2023} opens the possibility of black hole spectroscopy \cite{Berti2006}, whereby the QNM spectrum can be used to test deviations from the Kerr paradigm---including identifying signatures of regularity. Graybody factors (GBFs) \cite{Hawking1975,Page1976}, defined as the transmission probability $\Gamma_\ell(\omega)$ that a mode emitted near the horizon penetrates the potential barrier and escapes to infinity, determine the deviation of the Hawking radiation spectrum from a perfect blackbody \cite{Rincon2020}. They are intimately sensitive to the structure of the effective potential near its peak, which in turn is shaped by the higher-curvature corrections in QT gravity. Black hole shadows \cite{Cunha2018,Perlick2022,Synge1966}, the dark region bounded by the photon sphere's projection onto the observer's sky, have been imaged by the Event Horizon Telescope (EHT) for M87$^{*}$ \cite{EHT2019} and Sgr A$^{*}$ \cite{EHT2022}, providing a direct probe of strong-field gravity on horizon scales. The shadow radius $R_s$ is determined by the unstable circular null geodesics and thus encodes the metric structure in the photon sphere region.
	
	Remarkably, these three seemingly distinct observables are not independent. In the eikonal (large angular momentum $l \gg 1$) regime, a web of correspondences links them together. The Shadow-QNM correspondence, first identified by Jusufi \cite{Jusufi2020} and rooted in the fundamental connection between eikonal QNMs and unstable null geodesics established by Cardoso et al.\ \cite{Cardoso2009}, relates the shadow radius to the real part of the QNM frequency via $R_s \approx l/\omega_R$. The QNM-GBF correspondence, derived by Konoplya and Zhidenko within the WKB framework \cite{Konoplya2024,Konoplya2025}, expresses the graybody factor $\Gamma_\ell(\omega)$ analytically in terms of the fundamental QNM frequency $\omega_0$ and the first overtone $\omega_1$, with corrections up to $\mathcal{O}(l^{-2})$ beyond the eikonal limit. More recently, Pedrotti and Calz\`a \cite{Pedrotti2025} completed this ``trinity'' of correspondences by establishing the Shadow-GBF correspondence, which directly links the graybody factor to the shadow radius $R_s$ and the Lyapunov exponent $|\lambda|$ at the photon sphere. These correspondences are not merely of formal interest. They provide cross-checks between independent observational channels---a single measurement of the shadow radius, combined with the Lyapunov exponent, can in principle predict the entire QNM spectrum and the Hawking emission profile. Conversely, GW ringdown observations can constrain the shadow size, offering a multi-messenger test of gravity. However, the accuracy and domain of validity of these correspondences must be verified for each specific spacetime model, particularly when higher-curvature corrections modify the near-horizon geometry in nontrivial ways \cite{Bolokhov2026,Konoplya2024c,Skvortsova2025}.
	
	In this work, we systematically investigate three classes of observables---namely, the quasinormal modes, graybody factors, and black hole shadows---for a family of regular black holes in quasi-topological gravity, along with the correspondences among them. We present the corresponding numerical results and discuss the influence of the coupling constants in different schemes on these observables.
	
	The remainder of this paper is organized as follows. In Sec.~\ref{Quasi-topological gravity}, we outline the theoretical framework of quasi-topological gravity and introduce the five exact regular black hole solutions studied in this work. In Sec.~\ref{Correspondences}, we present the complete theoretical formalism for the scalar perturbations, the effective potential, the quasinormal modes, the graybody factors, and the black hole shadows, together with a full account of the correspondences among these three observables; we then present the numerical results and carry out a detailed analysis and discussion. Finally, in Sec.~\ref{Conclusion}, we summarize the paper and outline directions for future research.
	
	\section{Quasi-topological gravity}\label{Quasi-topological gravity}
	% ...
	In this section we first introduce the regular black hole \cite{Bueno2025} used in the present study. The line element of the $D$-dimensional spherically symmetric regular black hole spacetime is given by
	
	\begin{equation}
		ds^{2} = -N(r)^{2} f(r)\, dt^{2} + f^{-1}(r)\, dr^{2} + r^{2}\, d\Omega_{(D-2)}^{2},
		\label{eq:metric}
	\end{equation}
	
	where $d\Omega_{(D-2)}^{2}$ denotes the line element of the $(D-2)$-dimensional unit sphere. For $D = 4$, $d\Omega_{(2)}^{2} = d\theta^{2} + \sin^{2}\theta\, d\phi^{2}$, and the line element reduces to that of the four-dimensional spherically symmetric regular black hole.
	
	The quasi-topological theory yields the action
	
	\begin{equation}
		I_{\mathrm{QT}} = \frac{1}{16\pi G} \int d^{D}x \, \sqrt{|g|} \left[ R + \sum_{n=2}^{n_{\mathrm{max}}} \alpha_{n} \mathcal{Z}_{n} \right],
		\label{eq:action}
	\end{equation}
	
	where $\alpha_{n}$ are arbitrary coupling constants with dimensions of $[\mathrm{length}]^{2(n-1)}$. From the equations of motion of Eq.~\eqref{eq:action}, one obtains
	
	\begin{equation}
		\frac{dN}{dr} = 0, \qquad \frac{d}{dr} \left[ r^{D-1} h(\psi) \right] = 0,
		\label{eq:eom}
	\end{equation}
	
	where
	
	\begin{equation}
		h(\psi) \equiv \psi + \sum_{n=2}^{n_{\mathrm{max}}} \alpha_{n} \psi^{n}, \qquad h(\psi) = \frac{m}{r^{D-1}},
		\label{eq:hdef}
	\end{equation}
	
	with $m$ an integration constant directly related to the ADM mass $M$ of the black hole,
	
	\begin{equation}
		m \equiv \frac{16\pi G_{D} M}{(D-2) \Omega_{D-2}}, \qquad \Omega_{D-2} = \frac{(2\pi)^{\frac{D-1}{2}}}{\Gamma\left(\frac{D-1}{2}\right)}.
		\label{eq:mass}
	\end{equation}
	
	where $\Omega_{D-2}$ is the area of the unit $(D-2)$-sphere. Since $dN/dr=0$, $N$ is constant; rescaling the time coordinate, we set $N=1$. By choosing appropriate coupling parameters $\alpha_{n}$, one can ensure regular black hole solutions for which all curvature invariants remain finite everywhere.
	
	\begin{table*}[t]
		\caption{Regular black hole models used in this work. Models (a)--(e) are listed in Ref.~\cite{Bueno2025}; $r_0$ denotes the outer horizon radius, determined by $f(r_0) = 0$.}
		\label{tab:models}
		\renewcommand{\arraystretch}{1.6}
		\begin{ruledtabular}
			\large
			\begin{tabular}{c l l}
				Model & Metric function $f(r)$ & Constraints on $\alpha$ \\
				\hline
				\hypertarget{tab:a}{(a)} 
				& $1 - \frac{mr^2}{r^{D-1} + \alpha m}$ 
				& $0 \leq \frac{\alpha}{r_0^2} \leq \frac{D-3}{D-1}$ \\
				
				\hypertarget{tab:b}{(b)} 
				& $1 - \frac{mr^2}{\sqrt{r^{2(D-1)} + \alpha^2 m^2}}$ 
				& $0 \leq \frac{\alpha}{r_0^2} \leq \sqrt{\frac{D-3}{D-1}}$ \\
				
				\hypertarget{tab:c}{(c)} 
				& $1 - \frac{r^2}{\alpha} \left(1 - e^{-\alpha m / r^{D-1}}\right)$ 
				& $0 \leq \frac{\alpha}{r_0^2} \leq C(D) < 1$ \\
				
				\hypertarget{tab:d}{(d)} 
				& $1 - \frac{2mr^2}{r^{D-1} + \sqrt{r^{2(D-1)} + 4\alpha^2 m^2}}$ 
				& $0 \leq \frac{\alpha}{r_0^2} \leq \sqrt{\frac{D-3}{D+1}}$ \\
				
				\hypertarget{tab:e}{(e)} 
				& $1 - \frac{2mr^2}{r^{D-1} + 2\alpha m + \sqrt{r^{2(D-1)} + 4m\alpha r^{D-1}}}$ 
				& $0 \leq \frac{\alpha}{r_0^2} \leq \frac{D-3}{D+1}$ \\
			\end{tabular}
		\end{ruledtabular}
	\end{table*}
	
	Table~\ref{tab:models} lists the five regular black hole metric functions $f(r)$ obtained by Bueno et al.\ within the quasi-topological framework \cite{Bueno2025} by choosing different higher-order curvature coupling constants $\alpha_{n}$, together with their corresponding parameter constraints. All metrics share the same asymptotic structure $f(r\to\infty) \to 1 - m/r^{D-3}$ (i.e., recovering the $D$-dimensional Schwarzschild--Tangherlini form \cite{Tangherlini1963}), but they differ in their near-center behavior as $r \to 0$.
	
	Model \hyperlink{tab:a}{(a)} corresponds to the simplest geometric series $\alpha_{n} = \alpha^{n-1}$, which yields the higher-dimensional generalization of the Hayward regular black hole \cite{Hayward2006}; model \hyperlink{tab:b}{(b)}, whose coupling carries $\Gamma$-function weights, yields a Bardeen-like solution \cite{Bardeen1968}; model \hyperlink{tab:c}{(c)}, obtained by taking $\alpha_{n} = \alpha^{n-1}/n$, yields a Dymnikova-like solution with exponential corrections \cite{Dymnikova1992}; models \hyperlink{tab:d}{(d)}--\hyperlink{tab:e}{(e)} correspond respectively to higher-order generalizations under different summation schemes, such as retaining only odd-order terms and weighted series.
	
	The third column of the table gives the allowed range of the regularization parameter $\alpha$ in each model, expressed in the dimensionless form $\alpha/r_{0}^{2}$. The physical origin of these constraints lies in the horizon-existence condition: for a regular black hole to exist as a black hole, $f(r) = 0$ must admit at least one positive real root (the horizon). When $\alpha$ exceeds the critical value, the metric function is positive everywhere ($f(r) > 0$, $\forall r$), the horizon disappears, and the spacetime reduces to a horizonless but centrally regular compact object \cite{Dymnikova2020}.

\begin{figure}[t]
		\centering
		\includegraphics[width=1\columnwidth]{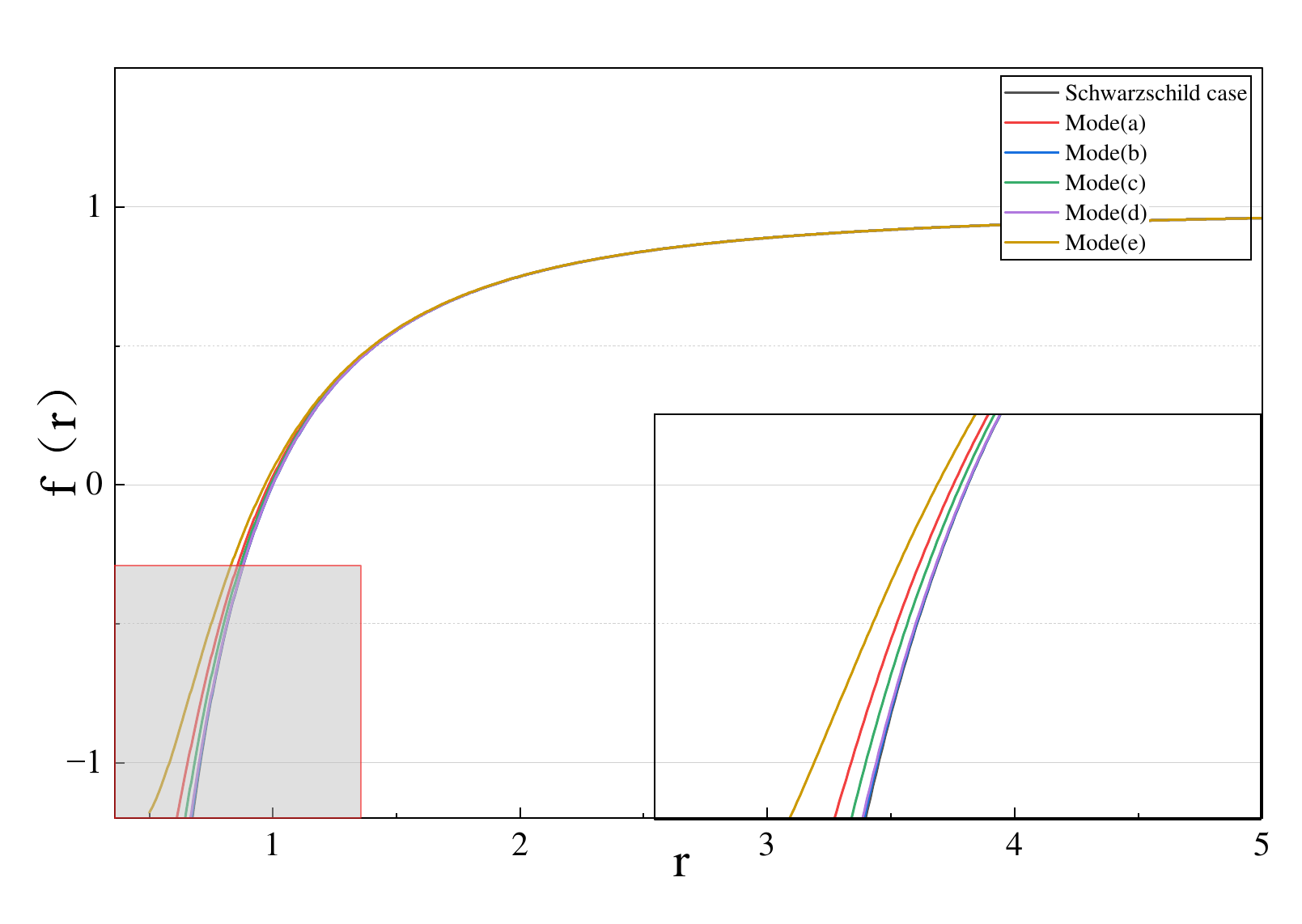}
		\caption{Metric functions $f(r)$ for models (a)--(e) with $\alpha = 0.03$, $m = 1$, and $D = 5$.}
		\label{fig:fig1}
	\end{figure}

	Figure~\ref{fig:fig1} shows the behavior of the metric function $f(r)$ as a function of the radial coordinate $r$ in five-dimensional ($D = 5$) spacetime, with the mass parameter $m = 1$ and the regularization parameter $\alpha = 0.03$. This metric arises from quasi-topological gravity, whose action contains higher-order curvature corrections of infinite order. In the large-scale limit $r \gg m^{1/(D-2)}$, $f(r) \to 1 - m/r^{D-3}$, which successfully recovers the asymptotic form of the $D$-dimensional Schwarzschild--Tangherlini solution \cite{Tangherlini1963}, consistent with the uniqueness of static spherically symmetric vacuum solutions required by Birkhoff's theorem \cite{Zegers2005}.
	
	The condition $f(r) = 0$ determines the location of the black hole horizon. The presence of the parameter $\alpha$ induces a small shift of the horizon location relative to the Schwarzschild case---consistent with the physical expectation that higher-order curvature corrections significantly modify the geometric structure only in the extremely high-curvature region approaching the Planck scale \cite{Addazi2022}.
	
	\section{Correspondences}\label{Correspondences}
	
	In this work we study the graybody factors (GBFs), the quasinormal modes (QNMs), and the black hole (BH) shadows in the presence of a massless scalar field $\Phi$, governed by the Klein--Gordon equation
	
	\begin{equation}
		\frac{1}{\sqrt{-g}} \partial_{\mu} \left( \sqrt{-g}\, g^{\mu\nu} \partial_{\nu} \Phi \right) = 0,
		\label{eq:kg}
	\end{equation}
	
	On the static, spherically symmetric background (Eq.~\eqref{eq:metric}) with $N=1$, the scalar field is decomposed into radial, angular, and temporal parts as
	
	\begin{equation}
		\Phi = \frac{\Psi_{\omega l}(r)}{r^{3/2}}\, Y_{l}(\Omega_{3})\, e^{-i\omega t},
		\label{eq:decomp}
	\end{equation}
	
	where $Y_{l}(\Omega_{3})$ are the hyperspherical harmonics on the unit three-sphere $S^{3}$, satisfying $\nabla^{2}_{S^{3}} Y_{l} = -l(l+2)\, Y_{l}$. Substituting Eq.~\eqref{eq:decomp} into Eq.~\eqref{eq:kg} and introducing the tortoise coordinate $dr_{*} \equiv dr/f(r)$, one obtains a Schr\"odinger-like equation for the radial function $\Psi_{\omega l}$,
	
	\begin{equation}
		\frac{d^{2}\Psi_{\omega l}}{dr_{*}^{2}} + \left[ \omega^{2} - V(r) \right] \Psi_{\omega l} = 0,
		\label{eq:schrodinger}
	\end{equation}
	
	where $\omega$ is the perturbation frequency, and the effective potential for the massless scalar field in $D=5$ dimensions is
	
	\begin{equation}
		V(r) = f(r) \left[ \frac{l(l+2)}{r^{2}} + \frac{3}{2}\frac{f'(r)}{r} + \frac{3}{4}\frac{f(r)}{r^{2}} \right],
		\label{eq:potential}
	\end{equation}
	
	with $f(r)$ given by Eq.~\eqref{eq:metric} and the prime denoting differentiation with respect to $r$.
	
	\begin{figure}[htbp]
		\centering
		\includegraphics[width=1\columnwidth]{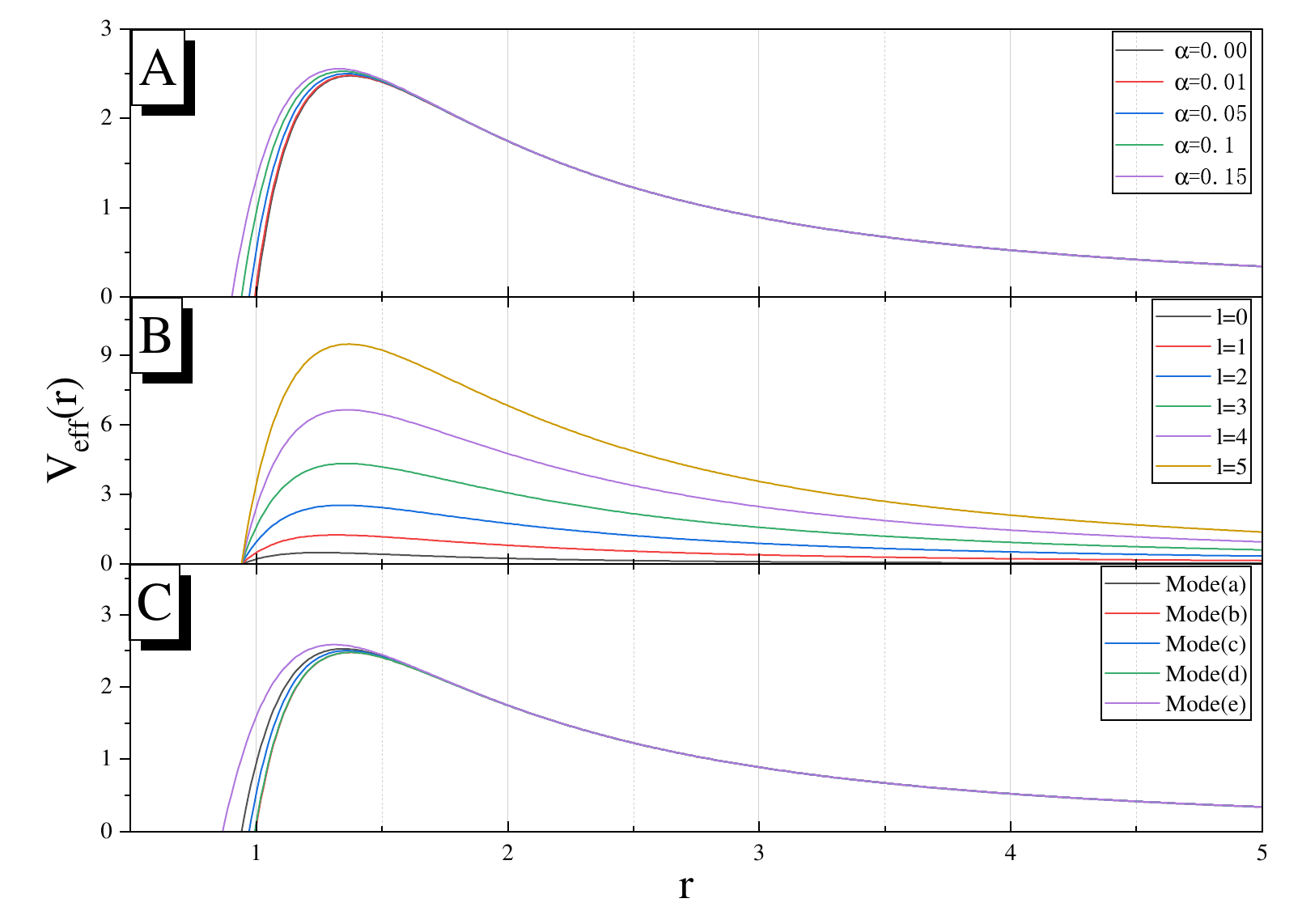}
		\caption{Effective potentials for the massless scalar field. (A) Model (a) with $l = 2$ and $D = 5$ for different values of $\alpha$. (B) Model (a) with $\alpha = 0.1$ and $D = 5$ for different values of $l$. (C) Comparison of the five regular models with the Schwarzschild metric for $l = 2$, $\alpha = 0.1$, and $D = 5$.}
		\label{fig:fig2}
	\end{figure}
	
	Figure~\ref{fig:fig2} shows the effective potential under different parameters. Panel (A) displays the effective potential of the scalar field for model \hyperlink{tab:a}{(a)} with $D = 5$ and $l = 2$ for different regularization parameters $\alpha$. It can be seen that all curves form a single-peak potential barrier. As $\alpha$ increases from $0$ to $0.15$, the peak of the barrier rises monotonically and its position shifts slightly toward smaller $r$---the reason being that $\alpha$ modifies the metric gradient near the horizon through $f'(r)$, rendering the near-region effective potential sensitive to the regularization parameter. Physically, a higher barrier implies a lower transmission probability for waves crossing the barrier. Meanwhile, according to the WKB formula~\eqref{eq:wkb-qnm}, an increase in $V_{0}$ leads to a larger real part $\omega_{R}$ of the QNM (faster oscillation) and a smaller imaginary part $|\omega_{I}|$ (slower decay) \cite{Konoplya2019}. In the far-field region of large $r$, the curves gradually converge, indicating that the higher-order curvature corrections are confined to the strong-gravity near region---a natural consequence of the fact that, in effective field theory, higher-derivative terms become significant only near the Planck scale \cite{Bueno2025,Konoplya2024c}.
	
	Panel (B) fixes $\alpha = 0.1$ and $D = 5$, and displays the influence of the orbital angular momentum $l = 0$--$5$ on the effective potential of the type-(a) black hole. The dominant term of $V(r)$ contains $l(l+2)/r^{2}$ (the centrifugal potential for $D = 5$), so increasing $l$ produces a threefold effect \cite{Konoplya2011,Toshmatov2015}:
		1. The barrier height increases sharply. This reflects the strong centrifugal repulsion experienced by high-angular-momentum modes---high-$l$ perturbations are harder to excite by the black hole, with correspondingly higher QNM excitation thresholds.
		2. The peak position of the barrier shifts outward. The peak location $r_{0}$ moves to larger $r$ as $l$ increases, and in the eikonal limit ($l \gg 1$), $r_{0}$ approaches the photon sphere radius $r_{\mathrm{ph}}$ \cite{Cardoso2009,Konoplya2022}.
		3. The special case $l = 0$ (s-wave). When $l = 0$ the centrifugal potential vanishes, so the barrier is lowest and flattest in shape. This means the s-wave is the mode most easily transmitted through the barrier, corresponding to the largest graybody factor and the fastest QNM decay.
	
	Panel (C) fixes $\alpha = 0.1$, $l = 2$, and $D = 5$, and compares the effective potentials of the five regular black hole models \hyperlink{tab:a}{(a)}--\hyperlink{tab:e}{(e)} with that of the Schwarzschild black hole ($\alpha = 0$). It can be found that the potential barriers of all five regular models are higher than the Schwarzschild baseline, and their peak positions are slightly shifted toward smaller $r$.

	\subsection{GBF,QNMs and BH shadow}
	\subsubsection{GBF}
	The graybody factor is defined as the transmission coefficient of the field through the black hole potential barrier, with boundary conditions requiring a purely ingoing wave at the event horizon and a superposition of ingoing and outgoing waves at infinity:
	
	\begin{equation}
		\begin{aligned}
			\Psi &= e^{-i\omega r_{*}} + R\, e^{i\omega r_{*}}, \qquad r_{*} \to +\infty, \\
			\Psi &= T\, e^{-i\omega r_{*}}, \qquad r_{*} \to -\infty.
		\end{aligned}
		\label{eq:bc}
	\end{equation}
	
	where the coefficients $T$ and $R$ correspond respectively to the transmission and reflection amplitudes, and their squared moduli give the transmission and reflection coefficients. For a given frequency $\omega$, the graybody factor is defined as
	
	\begin{equation}
		\Gamma_{\ell}(\omega) \equiv |T|^{2} = 1 - |R|^{2}.
		\label{eq:gbf}
	\end{equation}
	
	At present, the graybody factor can be obtained through the WKB approximation:
	
	\begin{equation}
		\Gamma_{\ell}(\omega) = \frac{1}{1 + e^{2\pi i \mathcal{K}}}.
		\label{eq:gbf-wkb}
	\end{equation}
	
	where $\mathcal{K}$ is a function of the frequency $\omega$, determined by the properties of the maximum of the effective potential. The quantity $\mathcal{K}$ satisfies the equation
	
	\begin{equation}
		\begin{aligned}
			\omega^{2} &\equiv V_{0} + \Lambda_{2}(\mathcal{K}^{2}) + \Lambda_{4}(\mathcal{K}^{2}) + \Lambda_{6}(\mathcal{K}^{2}) \\
			&\quad \pm i\mathcal{K}\sqrt{-2V_{0}^{\prime\prime}}\, \left(1 + \Lambda_{3}(\mathcal{K}^{2}) + \Lambda_{5}(\mathcal{K}^{2}) + \Lambda_{7}(\mathcal{K}^{2}) + \ldots \right).
		\end{aligned}
		\label{eq:wkb}
	\end{equation}
	
	where the explicit forms of $\Lambda_{i}$ for the second and third orders of the WKB method are given in Ref.~\cite{IyerWill1987}, those for the fourth to sixth orders in Ref.~\cite{Konoplya2003}, and those for the seventh to thirteenth orders in Ref.~\cite{Matyjasek2017}. Moreover, $\mathcal{K}$ satisfies
	
	\begin{equation}
		\mathcal{K} = n + \frac{1}{2}, \qquad n = 0, 1, 2, \ldots
		\label{eq:overtone}
	\end{equation}
	
	where $n$ is the overtone number.
	
	\begin{figure*}[t]
		\centering
		\includegraphics[width=\textwidth]{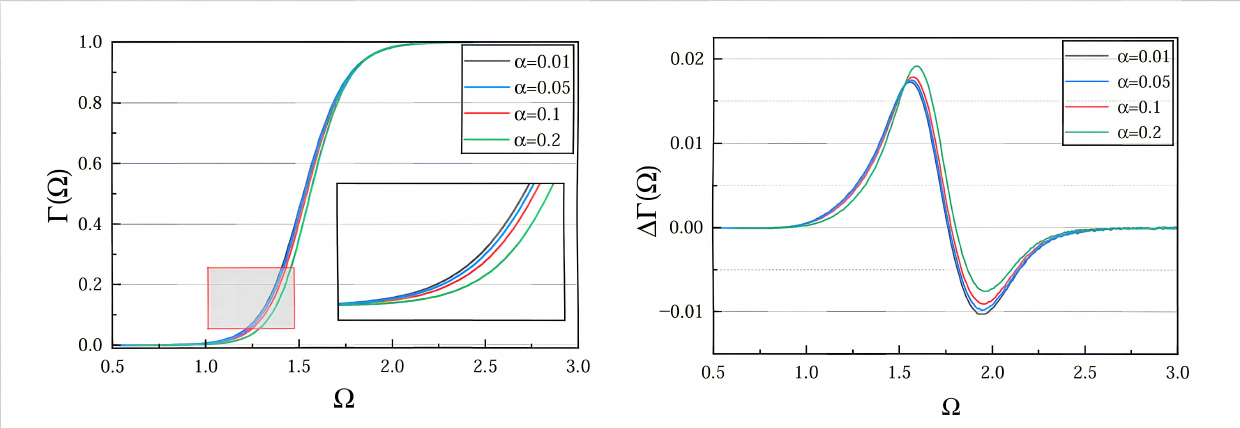}
		\caption{Difference between the GBFs obtained with GrayHawk and via the GBF--QNM correspondence for $D = 5$, $l = 2$, and different values of $\alpha$.}
		\label{fig:fig3}
	\end{figure*}
	
	In this work we use the publicly available package GrayHawk \cite{Calza2025} to perform the direct integration. GrayHawk is a Mathematica program specifically designed for computing the graybody factors of spherically symmetric black holes, supporting perturbations of scalar ($s = 0$), fermionic ($s = 1/2$), vector ($s = 1$), and tensor ($s = 2$) fields. By numerically inverting the tortoise coordinate and constructing the effective potential, it integrates the Schr\"odinger equation with NDSolve and performs a nonlinear fit in the far region to extract the transmission amplitude.
	
	Figure~\ref{fig:fig3} shows, for $D = 5$ and $l = 2$, the difference $\Delta\Gamma_{\ell}(\omega)$ between the GBF--QNMs correspondence (semi-analytical WKB) and the direct numerical integration results of GrayHawk for different regularization parameters $\alpha$. It is worth emphasizing that the Konoplya--Zhidenko correspondence holds exactly in the eikonal limit ($l \to \infty$), while $l = 2$ lies precisely in the low-mode region where the correspondence is least accurate; moreover, the correspondence was originally established only for four-dimensional spherically symmetric black holes, and its extension to $D = 5$ requires that the WKB correction coefficients $\Lambda_{i}$ at all orders retain their form in higher dimensions. Figure~\ref{fig:fig3} shows that even under the doubly unfavorable conditions of $l = 2$ and $D = 5$, $\Delta\Gamma_{\ell}(\omega)$ remains small in magnitude, indicating that the GBF--QNMs correspondence maintains its applicability in both the low-mode and higher-dimensional directions. From a physical point of view, the significance of this result lies in the fact that, once the QNM frequencies are known, the graybody factors can be obtained directly from the analytical formula, without the need to solve the scattering equation numerically for each frequency $\omega$. This is of practical value for computing quantities such as the Hawking radiation spectrum and the absorption cross section.

	\subsubsection{QNMs}
	The quasinormal modes of a black hole describe its characteristic oscillation modes after perturbation. The required boundary conditions are purely outgoing waves at infinity and purely ingoing waves at the event horizon:
	
	\begin{equation}
	\begin{aligned}
		\Psi &= A\, e^{+i\omega r_{*}}, \qquad r_{*} \to +\infty, \\
		\Psi &= B\, e^{-i\omega r_{*}}, \qquad r_{*} \to -\infty.
	\end{aligned}
	\end{equation}
	
	The QNM frequencies are generally written as
	
	\begin{equation}
		\omega_{\mathrm{QNM}} = \omega_{R} + i\,\omega_{I}.
		\label{eq:qnm-freq}
	\end{equation}
	
	where the real part $\omega_{R}$ corresponds to the oscillation frequency and the imaginary part $\omega_{I}$ to the damping. There are many methods for computing QNMs, including the WKB method \cite{Schutz1985}, the inverted potential method (IPM) \cite{Blome1984,Ferrari1984b,Ferrari1984a}, and the asymptotic iteration method (AIM) \cite{Cho2010,Cho2012,Ciftci2003}, among others. In this work we adopt the sixth-order WKB method.
	
	The 6th-order formalism of the WKB approximation has the formula
	
	\begin{equation}
		\frac{i\left(\omega_{\mathrm{QNM}}^{2} - V_{0}\right)}{\sqrt{-2V_{0}^{\prime\prime}}} - \sum_{k=2}^{6} \Lambda_{k} = n + \frac{1}{2}.
		\label{eq:wkb-qnm}
	\end{equation}
	
	where $V_{0}$ and $V_{0}^{\prime\prime}$ are respectively the maximum of the potential and its second derivative at the maximum, $n$ is the overtone number, and $k$ denotes the $k$-th-order correction. The formulae for $\Lambda_{2}$, $\Lambda_{3}$, $\Lambda_{4}$, $\Lambda_{5}$, and $\Lambda_{6}$ are given in Refs.~\cite{IyerWill1987,Konoplya2003}.

	\subsubsection{BH shadow}
	
	The black hole shadow refers to the spacetime region where photons are captured by the black hole's gravity and move along unstable null geodesics. Its boundary is defined by the light rays that are neither captured nor scattered by the black hole, but are temporarily trapped in the photon sphere.
	
	To compute the black hole shadow \cite{Tsupko2020}, we first define
	
	\begin{equation}
		F(r) \equiv \sqrt{\frac{r^2}{f(r)}},
		\label{eq:F}
	\end{equation}
	
	where $f(r)$ comes from Eq.~\eqref{eq:metric}. The photon sphere radius $r_{\mathrm{ph}}$ can be obtained by solving the equation
	
	\begin{equation}
		\frac{d}{dr}\left[ F^{2}(r_{\mathrm{ph}}) \right] = 0,
		\label{eq:photon}
	\end{equation}
	
	Substituting Eq.~\eqref{eq:F}, Eq.~\eqref{eq:photon} can be rewritten as
	
	\begin{equation}
		2\, f(r_{\mathrm{ph}}) - r_{\mathrm{ph}}\, f'(r_{\mathrm{ph}}) = 0,
		\label{eq:photon-cond}
	\end{equation}
	
	Finally, the shadow radius is given by the photon sphere radius $r_{\mathrm{ph}}$ via gravitational lensing \cite{Cunha2018,Tsupko2020},
	
	\begin{equation}
		R_{s} = \left. \sqrt{\frac{r^2}{f(r)}} \right|_{r = r_{\mathrm{ph}}}.
		\label{eq:shadow}
	\end{equation}
	
	\subsection{QNM-GBF correspondence}
	
	Under the eikonal approximation, the QNM--GBF correspondence, accounting for the second-order beyond-eikonal correction, is given by \cite{Konoplya2024}:
	
	\begin{equation}
		\resizebox{0.5\textwidth}{!}{%
			$\displaystyle
			\begin{aligned}
				i\mathcal{K} = &\frac{\omega^2 - {\text{Re}(\omega_0)}^2}{4\text{Re}(\omega_0)\text{Im}(\omega_0)} \left(1 + \frac{(\text{Re}(\omega_0) - \text{Re}(\omega_1))^2}{32{\text{Im}(\omega_0)}^2} - \frac{3\text{Im}(\omega_0) - \text{Im}(\omega_1)}{24\text{Im}(\omega_0)}\right) \\
				&-\frac{\text{Re}(\omega_0) - \text{Re}(\omega_1)}{16\text{Im}(\omega_0)} - \frac{(\omega^2 - {\text{Re}(\omega_0)}^2)^2}{{16\text{Re}(\omega_0)}^3\text{Im}(\omega_0)} \left(1 + \frac{\text{Re}(\omega_0)(\text{Re}(\omega_0) - \text{Re}(\omega_1))}{{4\text{Im}(\omega_0)}^2}\right) \\
				&+ \frac{(\omega^2 - {\text{Re}(\omega_0)}^2)^3}{{32\text{Re}(\omega_0)}^5\text{Im}(\omega_0)} \left(1 + \frac{\text{Re}(\omega_0)(\text{Re}(\omega_0) - \text{Re}(\omega_1))}{{4\text{Im}(\omega_0)}^2} + {\text{Re}(\omega_0)}^2 \times \right. \\
				&\quad \left. \left( \frac{(\text{Re}(\omega_0) - \text{Re}(\omega_1))^2}{{16\text{Im}(\omega_0)}^4} - \frac{3\text{Im}(\omega_0) - \text{Im}(\omega_1)}{12\text{Im}(\omega_0)} \right) \right) + \mathcal{O}\left(\frac{1}{l^3}\right)
			\end{aligned}
			$%
		}
	\end{equation}
	
	\subsection{Shadow-GBF correspondence}
	
	Recently, by combining the shadow--QNM connection with the QNM--GBF connection, researchers have obtained the shadow--GBF correspondence \cite{Pedrotti2025}:
	
	\begin{equation}
		\Gamma_{\ell}(\omega) = \left[ 1 + \exp\left( -2\pi R_{s}\, \frac{\omega^{2} - \left(l/R_{s}\right)^{2}}{2l|\lambda|} \right) \right]^{-1} + \mathcal{O}\left(l^{-1}\right).
		\label{eq:shadow-gbf}
	\end{equation}
	
	Here $R_{s}$ is the shadow radius, and $|\lambda|$ is the absolute value of the coordinate-time Lyapunov exponent associated with the instability of the photon sphere \cite{Cardoso2009}:
	
	\begin{equation}
		\lambda = \sqrt{ \frac{f(r_{\mathrm{ph}})}{2 r_{\mathrm{ph}}^{2}} \left[ 2f(r_{\mathrm{ph}}) - r_{\mathrm{ph}}^{2} f''(r_{\mathrm{ph}}) \right] }.
		\label{eq:lyapunov}
	\end{equation}
	
	This correspondence holds in the eikonal limit and the WKB approximation.
	
	We note that the strict eikonal parameter in the above formula is not $l$, but rather the effective angular momentum after the Langer correction (to be distinguished from the WKB parameter $\mathcal{K}$ introduced above),
	
	\begin{equation}
		\kappa = \sqrt{l(l+D-3)} \simeq l + \frac{D-3}{2},
		\label{eq:kappa}
	\end{equation}
	
	where $l(l+D-3)$ is the eigenvalue of the Laplacian of the scalar field on $S^{D-2}$. The physical basis for this replacement is that the next-to-leading-order formula of the shadow--QNM correspondence in higher dimensions is $\omega_{R} = R_{s}^{-1} \left[ l + \frac{D-3}{2} + \mathcal{O}(l^{-1}) \right]$ \cite{CuadrosMelgar2020}, therefore, upgrading $\omega_{R} = l/R_{s}$ to $\omega_{R} = \kappa/R_{s}$ exactly absorbs this next-to-leading term. On the other hand, in the eikonal limit, the leading order of $\omega_{I} = -(n + 1/2)|\lambda|$ is dimension-independent, so the Lyapunov exponent requires no correction. Accordingly, for $D = 5$ one has $\kappa = l + 1$, and the shadow--GBF correspondence is modified to
	
	\begin{equation}
		\Gamma_{l}(\omega) = \frac{1}{1 + \exp\left( -\pi R_{s}\, \frac{\omega^{2} - \kappa^{2}/R_{s}^{2}}{\kappa |\lambda|} \right)}.
		\label{eq:shadow-gbf-mod}
	\end{equation}
	
	This modification does not change the structure of the formula, but merely replaces $l$ with $\kappa$; nevertheless, it improves the accuracy for small $l$.
	
	\begin{figure}[htbp]
		\centering
		\includegraphics[width=1.1\columnwidth]{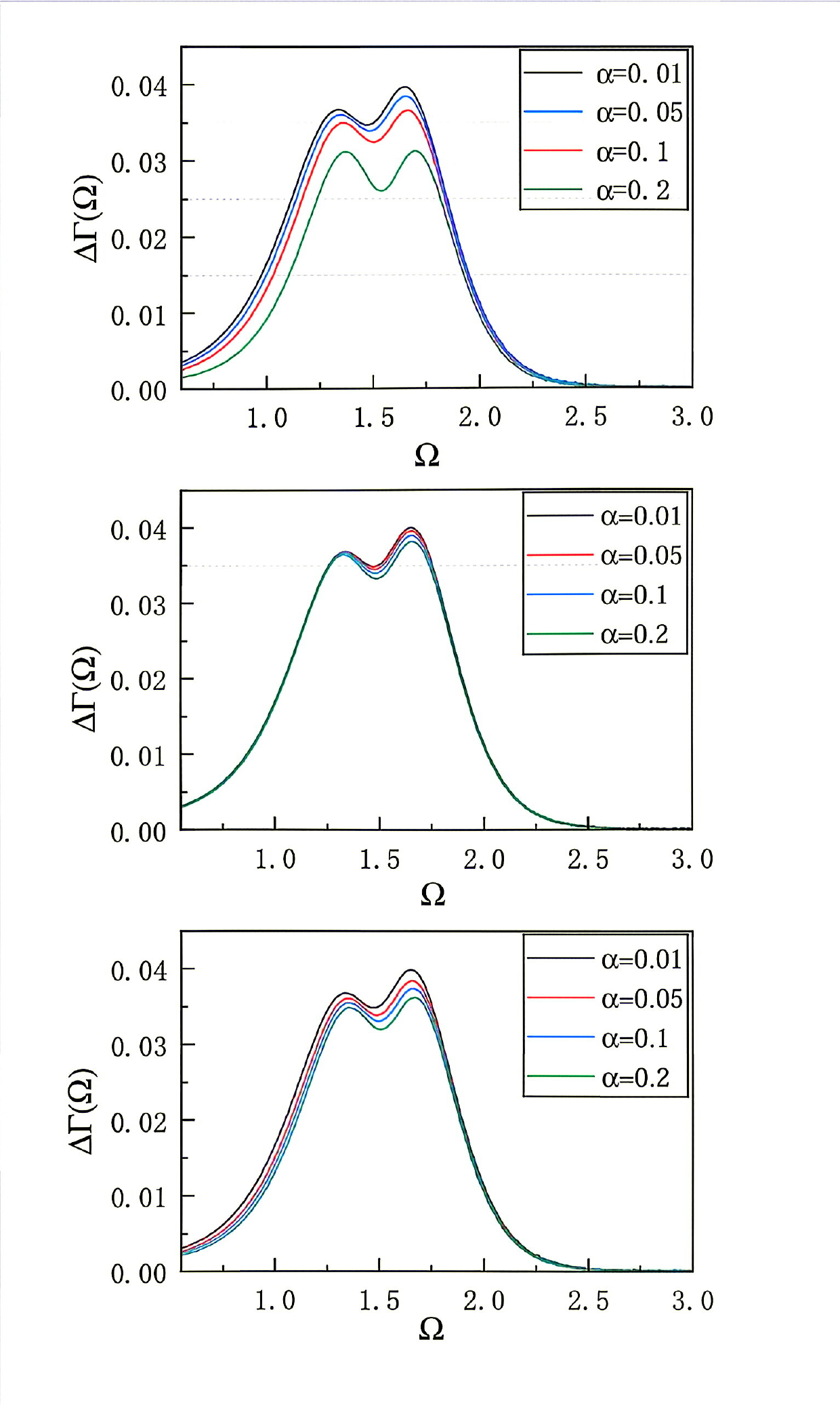}
		\caption{Difference between the shadow--GBF correspondence and the direct numerical integration for $D = 5$ and $l = 2$.}
		\label{fig:fig4}
	\end{figure}
	
	The shadow--GBF correspondence \cite{Pedrotti2025} is obtained by concatenating two correspondences that hold exactly in the eikonal limit: the first is the shadow--QNM correspondence \cite{Cardoso2009,CuadrosMelgar2020,Jusufi2020}, which relates the black hole shadow radius to the eikonal QNM frequencies; the second is the GBF--QNMs correspondence \cite{Konoplya2024}, which uses the WKB expansion to express the graybody factor as a function of the fundamental mode and the first overtone. Therefore, the shadow--GBF relation is exact as $l \to \infty$ and approximate for finite $l$, with its error originating from the asymptotic nature of the WKB series.
	
	It is worth noting that \cite{Pedrotti2025} verified the shadow--GBF correspondence only in four dimensions and mainly for high multipole numbers ($l \gtrsim 10$). In this work, we take the lowest multipole number $l = 2$ in $D = 5$, thereby testing the applicability of this correspondence in both the higher-dimensional and low-mode directions. Low modes give the dominant contribution to the Hawking radiation spectrum (the geometric potential grows with $l$), and are therefore particularly important for physical applications. We regard the GrayHawk numerical results as the ``exact'' benchmark, while the shadow--GBF correspondence provides a WKB semi-analytical approximation. Their difference $\Delta\Gamma_{l}(\omega)$ thus quantifies the systematic error of the correspondence at finite $l$, rather than a numerical error.
	
	Figure~\ref{fig:fig4} shows that, in $D = 5$, the error of the shadow--GBF correspondence for the low mode ($l = 2$) is already of the same order of magnitude as that of the GBF--QNMs correspondence (Fig.~\ref{fig:fig3}). Since the shadow--GBF relation is constructed precisely with the GBF--QNMs relation as the intermediate link, this result suggests that the error originates mainly from the WKB approximation in the GBF--QNMs link, rather than from the shadow--QNM link, thereby supporting the validity of the shadow--GBF correspondence in higher dimensions at the next-to-leading order.
	
	\section{Conclusion}\label{Conclusion}
	% ...
	In this work, within the framework of quasi-topological gravity, we have carried out a systematic study of three classes of observables---the shadow, the quasinormal modes, and the graybody factors---for a family of regular black holes, together with their correspondences.
	
	In $D \geq 5$ dimensions, by choosing different summation schemes for the infinitely many higher-order curvature correction coupling constants $\alpha_{n}$, quasi-topological gravity can systematically yield five types of regular black hole solutions \hyperlink{tab:a}{(a)}--\hyperlink{tab:e}{(e)}, corresponding respectively to the higher-dimensional generalizations of the Hayward, Bardeen, and Dymnikova solutions, as well as several new solutions. All these solutions possess a central de Sitter core, are free of singularities, and asymptotically recover the Schwarzschild--Tangherlini form; moreover, as pure-gravity vacuum solutions they require no matter fields, theoretically confirming that the pure-gravity mechanism alone can achieve singularity resolution. The regularization parameter $\alpha$ modifies the metric structure only in the near-center (strong-gravity) region, and is constrained by the horizon-existence condition; when $\alpha$ exceeds the critical value, the horizon disappears and the spacetime reduces to a regular compact object without an event horizon.
	
	The effective potential analysis shows that, under scalar-field perturbations, all five regular metrics form a single-peak potential barrier, generally higher than the Schwarzschild baseline. Increasing $\alpha$ raises the barrier peak and slightly shifts its position inward, corresponding to faster QNM oscillation and slower decay; increasing the angular momentum $l$ sharply raises the barrier and shifts its peak outward, with the peak position approaching the photon sphere radius in the eikonal limit, while the s-wave ($l = 0$) has the lowest, most easily penetrated barrier and hence the largest graybody factor. The higher-order curvature corrections are confined to the strong-gravity near region, consistent with the natural expectation that, in effective field theory, higher-derivative terms become significant only near the Planck scale.
	
	Regarding the verification of the correspondences, this work focuses on the applicability of the GBF--QNM and shadow--GBF correspondences in two directions: the low mode ($l = 2$) and higher dimensions ($D = 5$). The results show that, even under the doubly unfavorable conditions, the difference between the GBF--QNM correspondence and the GrayHawk direct numerical integration results remains small in magnitude, confirming its applicability; the physical significance lies in the fact that, once the QNM frequencies are known, the graybody factors can be obtained analytically, avoiding the point-by-point numerical solution of the scattering equation. Furthermore, after the Langer correction (replacing $l$ by the effective angular momentum $\kappa = \sqrt{l(l+D-3)} \simeq l + (D-3)/2$), the error of the shadow--GBF correspondence at $D = 5$, $l = 2$ is of the same order as that of the GBF--QNM correspondence, indicating that the shadow--QNM link introduces no additional dominant error, thereby supporting the validity of the shadow--GBF correspondence in higher dimensions at the next-to-leading order. This result closes the ``shadow--quasinormal modes--graybody factors'' trinity of correspondences, providing a viable route to approximately predict the eikonal QNM frequencies and the Hawking radiation profile from the shadow as a single observable, and thereby to test gravitational theories through multi-messenger observations.
	
	The main innovations of this work can be summarized in three points. First, to the best of our knowledge, this is the first work to test the shadow--GBF correspondence on regular black holes in $D \geq 5$-dimensional quasi-topological gravity: this correspondence was previously verified only in four dimensions and mainly for high multipole numbers ($l \gtrsim 10$) of the Schwarzschild, Bardeen, and Hayward black holes \cite{Pedrotti2025}; we extend it to higher-dimensional regular black holes, thereby broadening its range of applicability in the dimensionality direction. Second, the verification is carried out at the low mode $l = 2$: the shadow--GBF correspondence is built upon the eikonal limit, being exact as $l \to \infty$ and merely approximate for finite $l$; the fact that the error remains small in the $l = 2$ region constitutes a test of the robustness of this correspondence. Third, we propose the Langer correction: replacing $l$ in the formula by the effective angular momentum $\kappa = \sqrt{l(l+D-3)} \simeq l + (D-3)/2$ ($\kappa = l + 1$ for $D = 5$), thereby absorbing the $(D-3)/2$ next-to-leading term of the shadow--QNM correspondence in higher dimensions \cite{CuadrosMelgar2020}; this modification does not change the structure of the formula, yet reduces the error of the shadow--GBF correspondence at low modes to a level comparable with that of the GBF--QNM correspondence.
	
	The limitation of this work is that the verification is restricted to the static spherically symmetric case, scalar-field perturbations, and $D = 5$. In the future, this framework can be extended to rotating black holes, higher dimensions ($D > 5$), and gravitational tensor perturbations ($s = 2$), and applied to the concrete computation of the Hawking evaporation spectrum and the absorption cross section, in conjunction with observations such as EHT and LISA.

	\begin{acknowledgments}
		Your acknowledgments here.
	\end{acknowledgments}
	
	\bibliographystyle{apsrev4-2}
	\bibliography{ref}   
	
\end{document}